\documentclass[letterpaper,twocolumn,superscriptaddress,floatfix]{revtex4}
\usepackage[latin1]{inputenc}
\usepackage{bm}
\usepackage{multirow,amssymb,amsbsy,amsmath}
\usepackage{graphicx}
\usepackage{verbatim}
\usepackage{mathrsfs}
\usepackage{color}
\makeatletter
\usepackage{pifont}
\usepackage{makecell}
\newcommand{\tabincell}[2]{\begin{tabular}{@{}#1@{}}#2\end{tabular}}

\makeatother

\begin{document}
	\renewcommand{\figurename}{Fig.}
	
	\title{Coherent control of an ultrabright single spin in hexagonal boron nitride at room temperature}
	
	\author{Nai-Jie Guo}
	\thanks{These authors contributed equally to this work.}
	\affiliation{CAS Key Laboratory of Quantum Information, University of Science and Technology of China, Hefei, Anhui 230026, China.}
	\affiliation{CAS Center For Excellence in Quantum Information and Quantum Physics, University of Science and Technology of China, Hefei, Anhui 230026, China.}
	\affiliation{Hefei National Laboratory, University of Science and Technology of China, Hefei 230088, China}
	\author{Song Li}
	\thanks{These authors contributed equally to this work.}
	\affiliation{Wigner Research Centre for Physics, H-1121 Budapest, Hungary}
	\author{Wei Liu}
	\thanks{These authors contributed equally to this work.}
	\author{Yuan-Ze Yang}
	\thanks{These authors contributed equally to this work.}
	\author{Xiao-Dong Zeng}
	\thanks{These authors contributed equally to this work.}
	\author{Shang Yu}
	\author{Yu Meng}
	\author{Zhi-Peng Li}
	\author{Zhao-An Wang}
	\author{Lin-Ke Xie}
	\affiliation{CAS Key Laboratory of Quantum Information, University of Science and Technology of China, Hefei, Anhui 230026, China.}
	\affiliation{CAS Center For Excellence in Quantum Information and Quantum Physics, University of Science and Technology of China, Hefei, Anhui 230026, China.}
	\affiliation{Hefei National Laboratory, University of Science and Technology of China, Hefei 230088, China}
	\author{Rong-Chun Ge}
	\affiliation{College of Physics, Sichuan University, Chengdu China, 610064}
	\author{Jun-Feng Wang}
	\affiliation{CAS Key Laboratory of Quantum Information, University of Science and Technology of China, Hefei, Anhui 230026, China.}
	\affiliation{CAS Center For Excellence in Quantum Information and Quantum Physics, University of Science and Technology of China, Hefei, Anhui 230026, China.}
	\affiliation{Hefei National Laboratory, University of Science and Technology of China, Hefei 230088, China}
	\affiliation{College of Physics, Sichuan University, Chengdu China, 610064}
	\author{Qiang Li}
	\author{Jin-Shi Xu}
	\author{Yi-Tao Wang}
	\email{yitao@ustc.edu.cn}
	\author{Jian-Shun Tang}
	\email{tjs@ustc.edu.cn}
	\affiliation{CAS Key Laboratory of Quantum Information, University of Science and Technology of China, Hefei, Anhui 230026, China.}
	\affiliation{CAS Center For Excellence in Quantum Information and Quantum Physics, University of Science and Technology of China, Hefei, Anhui 230026, China.}
	\affiliation{Hefei National Laboratory, University of Science and Technology of China, Hefei 230088, China}
	\author{Adam Gali}
	\email{gali.adam@wigner.hu}
	\affiliation{Wigner Research Centre for Physics, H-1121 Budapest, Hungary}
	\affiliation{Department of Atomic Physics, Institute of Physics, Budapest University of Technology and Economics, H-1111 Budapest, Hungary}
	\author{Chuan-Feng Li}
	\email{cfli@ustc.edu.cn}
	\author{Guang-Can Guo}
	\affiliation{CAS Key Laboratory of Quantum Information, University of Science and Technology of China, Hefei, Anhui 230026, China.}
	\affiliation{CAS Center For Excellence in Quantum Information and Quantum Physics, University of Science and Technology of China, Hefei, Anhui 230026, China.}
	\affiliation{Hefei National Laboratory, University of Science and Technology of China, Hefei 230088, China}

	\date{\today}
	
	\begin{abstract}
		\textbf{Abstract}
		
		\textbf{Hexagonal boron nitride (hBN) is a remarkable two-dimensional (2D) material that hosts solid-state spins and has great potential to be used in quantum information applications, including quantum networks. However, in this application, both the optical and spin properties are crucial for single spins but have not yet been discovered simultaneously for hBN spins. Here, we realize an efficient method for arraying and isolating the single defects of hBN and use this method to discover a new spin defect with a high probability of 85\%. This single defect exhibits outstanding optical properties and an optically controllable spin, as indicated by the observed significant Rabi oscillation and Hahn echo experiments at room temperature. First principles calculations indicate that complexes of carbon and oxygen dopants may be the origin of the single spin defects. This provides a possibility for further addressing spins that can be optically controlled.}
		
	\end{abstract}
	

	\maketitle
	\textbf{Introduction}
	
	Solid-state spin defects play a crucial role in quantum information applications \cite{Wolfowicz2021}, especially color-center defects that have electronic spins that can be optically initialized and provide readouts, such as the nitrogen-vacancy (NV) center in diamond \cite{Doherty2013} and the divacancy defect in silicon carbide \cite{Lohrmann2017}. The coherent control of these spin defects is possible even under ambient conditions (e.g., room temperature and normal atmosphere) \cite{Awschalom2018}. Hence, they are practical resources in the construction of room-temperature spintronic quantum devices \cite{Atature2018}. Furthermore, an individual solid-state spin defect exhibits the characteristics of single-photon emission and nanoscale spatial resolution, and it is critical for various quantum-information applications, such as quantum photon source \cite{Aharonovich2016}, nanoscale quantum sensors \cite{Schirhagl2014} and quantum networks \cite{Bernien2013}. Especially in quantum networks, photon-spin entanglement and coupling play crucial roles. To achieve these aims, both the optical and spin performances are needed for this single spin.
	
	Recently, hexagonal boron nitride (hBN) has attracted considerable attention as a wide-band-gap van der Waals material that can host color-center spin defects \cite{Caldwell2019}. The spin defects in hBN benefit from the unique physical properties of van der Waals materials, e.g., flexible mechanical properties and natural two-dimensional characteristics; hence, they possess multiple modulation mechanisms and have great potential to be applied in the fabrication of two-dimensional quantum devices and integrated quantum nanodevices \cite{Grosso2017,Liu2020,tetienne2021,Liu2016}.
	
	The discovered spin defects in hBN include the negatively charged boron vacancy (V$_\text{B}^-$) defect \cite{gottscholl2020Initialization,kianinia2020generation,gao2021femtosecond,murzakhanov2021creation,guo2021generation,ivady2020ab,reimers2020photoluminescence,mathur2021excited,baber2021excited,yu2021excited,mu2022excited,liu2021temperature,gottscholl2021spin} and several carbon-related defects \cite{chejanovsky2021,mendelson2021,stern2021room}. At room temperature, the spin state of the V$_\text{B}^-$ defect exhibits obvious contrast under optically detected magnetic resonance (ODMR) \cite{gao2021} and Rabi oscillation with microwave (MW) resonance \cite{gottscholl2021,liu2021rabi}. This indicates that the optical initialization/readout and coherent control of V$_\text{B}^-$ in hBN are possible under the ambient conditions, but this is only possible for the ensemble. The isolation of a single V$_\text{B}^-$ defect is still challenging due to its poor optical properties. Compared to other color-center defects, V$_\text{B}^-$ defects in hBN display extremely low quantum efficiency for the optical transition. This low-quantum-efficiency characteristic makes it difficult to detect a single V$_\text{B}^-$ defect, although many studies have succeeded in effectively enhancing the photoluminescence (PL) of the V$_\text{B}^-$ defect \cite{froch2021coupling,mendelson2021coupling,gao2021}. In addition, the PL of V$_\text{B}^-$ always exhibits a broad spectrum with a considerable phonon sideband (PSB), even in the liquid-helium temperature region, and the zero-phonon-line (ZPL) does not appear. For some other defects in hBN, single spins with ODMR are found, but the probability is too low ($ \sim 5\% $ \cite{stern2021room}). Considering the low probability of detecting the Rabi signal ($ 10.5\% $, see data in Table II), it is difficult to observe the obvious coherent control of a single spin. In addition, other optical properties of these defects, for example, the Debye-Waller (DW) factor, still have room for improvement. These problems hinder the room-temperature quantum control of a single spin with good optical performance in hBN, which is crucial for quantum information applications based on solid-state spin defects.

	Here, we experimentally demonstrate an effective method for generating the color-center array of hBN on a gold-film MW waveguide and enhance the probability of finding a single spin with an ODMR ($ 85\% $) by 21-fold. Based on this progress, we realize the room-temperature coherent control of a single spin defect in hBN. We discover a new single spin in this array, which has both outstanding optical and spin properties that can potentially be used in quantum information applications. This defect has $ g^{(2)}(0)=0.25 $, an emission rate of 25 MHz (corrected), a DW factor of 0.8, an ODMR contrast of $ 2\% $, a linewidth of 22 MHz, clear Rabi oscillation, and $ T_1=16.17 $ $ \mu $s and $ T_2=2.45 $ $ \mu $s (from Hahn-echo experiment). The values of $ T_1 $ and $ T_2 $ reach the level of the V$_\text{B}^-$ defect. We also carried out \textit{ab initio} simulations, and the results imply that this new defect may be a complex of carbon and oxygen dopants.
	
	\begin{figure*}
		\centering
		\includegraphics[width=0.75\textwidth]{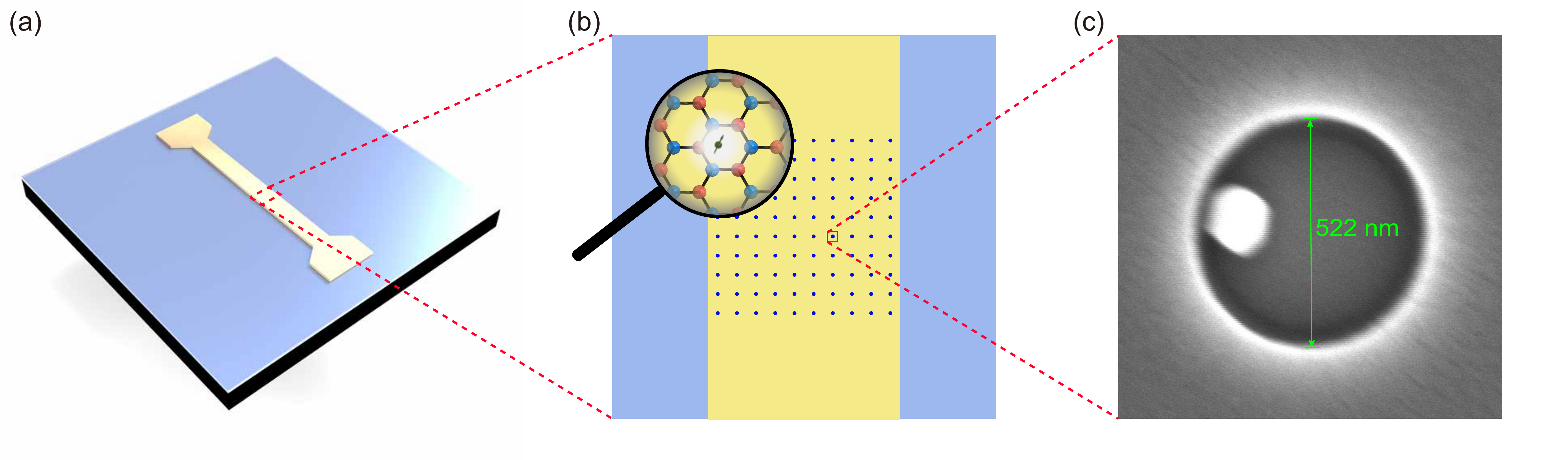}
		\caption{\label{Fig1} \textbf{Schematic of hBN arrays on the gold film MW waveguide.} (a) A schematic of a gold film MW waveguide on silicon wafer generating a uniform and stable local MW field. (b) A schematic of hBN arrays by dispersing the hBN suspension into the holes formed by electron beam lithography. Inset: Schematic of the atomic structure of hBN. (c) Scanning electron microscope image of an isolated hBN flake in one hole of the array before the PMMA resist was dissolved.}
	\end{figure*}
	
	\textbf{Results}

	\textbf{Color-center array.} To generate the color-center array of hBN, we use electron beam lithography to produce hole-array patterning on a silicon dioxide substrate coated with polymethyl methacrylate (PMMA) resist, and suspend the dispersed hBN ultrafine powder in the holes by capillary force \cite{Preu2021assembly,Malaquin2007controlled,Cui2004Integration}. The PMMA resist is subsequently dissolved, leaving the hBN-flake array on the gold film MW waveguide, as shown in Fig. 1. The gold film MW waveguide fabricated by photolithography under the spin-defect array can provide a uniform and stable local MW field and enhance the quantum efficiency and reflection efficiency of the spin defect \cite{gao2021,yu2021excited,mathur2021excited}. The scanning electron microscope (SEM) image of the hBN nanoflake in a single hole of the array is shown in Fig. 1c, showing that the single hBN flake can be isolated effectively. The hBN flakes originally host optically active defects \cite{Cassabois2016Hexagonal,Weng2017Tuning,Bourrellier2016Bright,Chen2021solvent,Jungwirth2016temperature,Tran2016Robust}, and this process increases the possibility of isolating a single nanoflake with a single spin defect instead of a cluster of flakes. See the experimental section for details on the sample preparation process.
	
	\begin{figure*}
		\centering
		\includegraphics[width=0.9\textwidth]{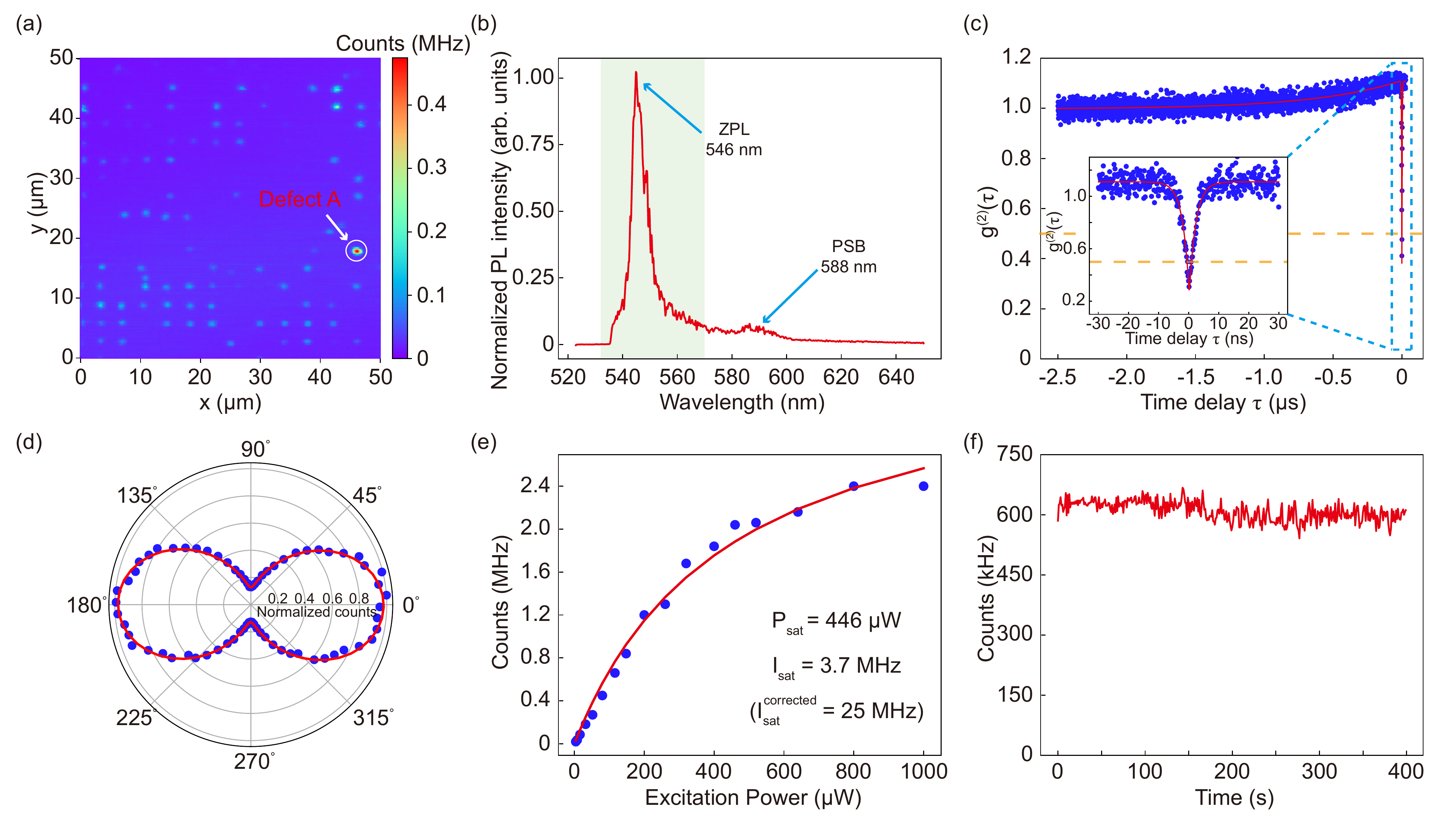}
		\caption{\label{Fig2} \textbf{Optical properties of the single spin in hBN.} (a) 50$ \times $50-$ \mu $m$ ^2 $ confocal PL map of hBN arrays with 100-$ \mu $W laser excitation at 532 nm. Our measurements correspond to the ultrabright single spin of Defect A circled in the image. (b) Room-temperature photoluminescence spectrum of Defect A under 532-nm laser excitation and with a 532-nm longpass filter. The zero-phonon line (ZPL) is indicated by green bands and the phonon sideband (PSB) is located at $ \sim $ 588 nm. (c) Second-order autocorrelation measurement g$ ^{(2)}(\tau) $ of Defect A, which is measured at 100-$ \mu $W laser power. The blue dots are the experimental data, the red curve is a fit based on the three-state model, and the yellow line represents $ g^{(2)}(\tau) =0.5$. Inset: A zoom of the same measurement for short timescales indicates the fitted $ g^{(2)}(0) =0.25\pm 0.02$. (d) Emission (blue dots) polarization behaviors of Defect A. The red curve is fitted using a $ \cos ^{2}\theta $ function. (e) Saturation curve of Defect A under 532-nm laser excitation. "Corrected" means that the saturated intensity is corrected by the photon collecting and detecting efficiencies. See Supplementary Note 1. (f) Time-dependence of photoluminescence counts of Defect A, showing stable emission from Defect A.}
	\end{figure*}
	
	The confocal PL map of the hBN array is shown in Fig. 2a, which is scanned by a home-built confocal microscope. Defect A is obviously brighter than the other defects, and we perform the characterizations mainly on Defect A, as reported in the following.
	
	\textbf{Optical properties.} The room-temperature PL spectrum of Defect A is shown in Fig. 2b, with a ZPL at $ \sim $ 546 nm (which is coincident with the theoretical prediction; See Supplementary Fig. 6) and full width at half maximum (FWHM) $ =6.5 $ nm. A weak PSB peak is located at $ \sim $ 588 nm corresponding to the 162-meV phonon. The DW factor, defined as the ratio of the integrated intensity of the ZPL to the total intensity of the emission, is derived as $ \sim $ 0.8. This DW factor indicates weak electron-phonon coupling and is one of the highest values of the reported quantum emitters in hBN at room temperature \cite{Tran2016Quantum}. The emission polarization behaviors of Defect A are shown in Fig. 2d, where the degree of polarization (DOP) equals to $ \sim $ 0.8, which is consistent with the characteristic of a single linearly polarized dipole. The emission stability of Defect A is shown in Fig. 2f, exhibiting stable fluorescence emission with no significant blinking over 400 s, and we also do not observe the bleaching effect of this defect during data measurement.
	
	We also perform the second-order autocorrelation measurement for the photons emitted from Defect A at 100-$ \mu $W laser power. The result is shown in Fig. 2c where $ g^{(2)}(0) =0.25\pm 0.02 <0.5$, indicating that Defect A acts as a single-photon emitter. The PL counts of Defect A are presented in Fig. 2e. The emission rate under continuous excitation and excitation power at saturation are $ I_{\text{sat}}=3.7\pm 0.20 $ MHz and $ P_{\text{sat}}=446\pm 54 $ $\mu$W, respectively. When taking the collection and detection efficiency into consideration, the corrected saturation emission rate can reach 25 MHz (See Supplementary Note 1 for details). The above optical properties of a high DW factor, high DOP and high optical stability reveal that the trapping-electron state in Defect A is hardly disturbed by the surroundings, and the optical properties of low $ g^{(2)}(0) $ and high brightness suggest that Defect A is a single optically active defect, possibly with high quantum efficiency.
	
	\begin{figure*}
		\centering
		\includegraphics[width=0.9\textwidth]{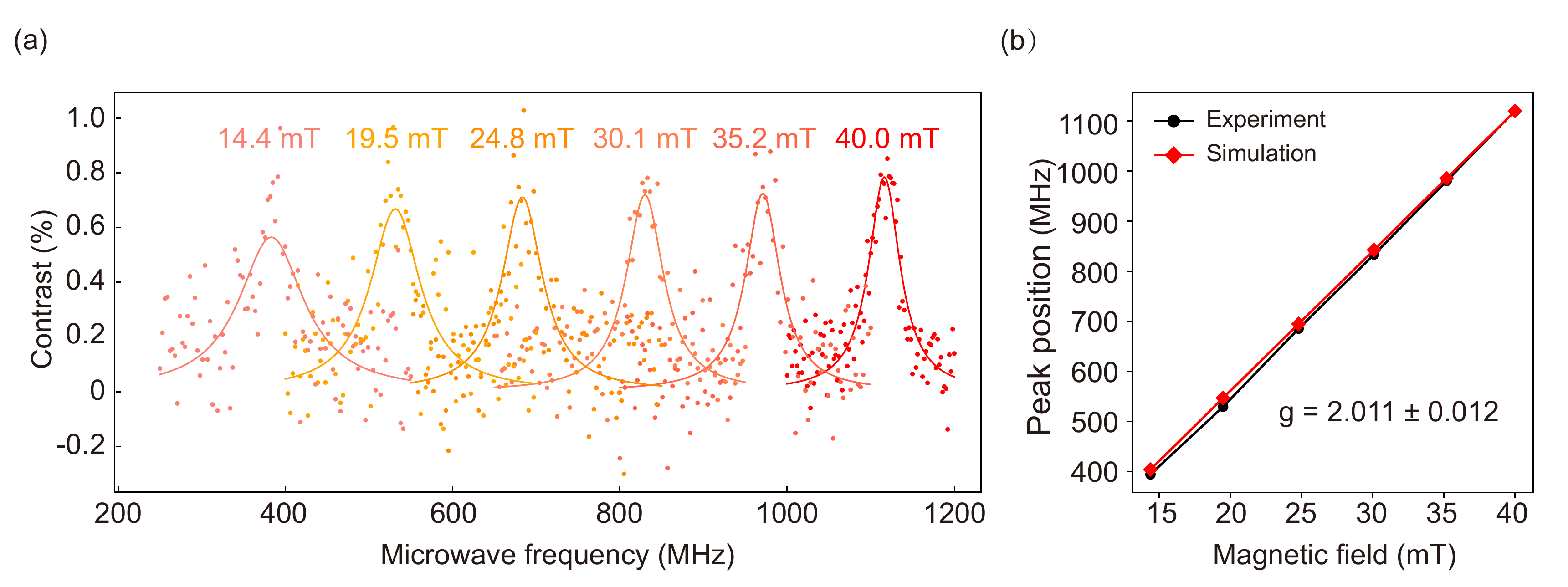}
		\caption{\label{Fig3} \textbf{Optically detected electron spin resonance of Defect A at room temperature under 100-$ \mu $W laser excitation and 25-dBm MW power.} (a) ODMR spectra of Defect A measured at different external magnetic field. (b) Dependence of ODMR resonance frequencies on the magnetic field (\textbf{B} $ || $ \textbf{c}). The black dots are experimental results extracted from (a), based on which we fit the $ g $ factor of Defect A to be $ 2.011 \pm  0.012 $. The red dots are \textit{ab initio} theoretical predictions.}
	\end{figure*}
	
	\textbf{ODMR measurements.} To determine whether Defect A has a spin that can be optically initialized and detected, we perform ODMR measurements under 100-$\mu$W, 532-nm laser excitation and 25-dBm MW power. The results are shown in Fig. 3. An electromagnet is used to provide the variable and controllable external magnetic field parallel to the hexagonal c axis (\textbf{B} $ || $ \textbf{c}) of hBN. The ODMR spectra with the single positive peak of Defect A at the different magnetic fields are displayed in Fig. 3a. As the magnetic field intensity decreases, the ODMR contrast decreases slightly and the linewidth increases slightly (we note that this may be caused by the decreasing signal-to-noise ratio, which impacts the fitting results of these ODMR spectra; in the following experiment, we will show that the linewidth is only dependent on the laser and MW powers.). The ODMR positive peak exhibits a contrast of 0.8\% and a linewidth of 37 MHz under the current laser and MW powers for the 40-mT data, which has the highest signal-to-noise ratio. The relation of the ODMR peak position and the magnetic-field strength is shown in Fig. 3b, which is coincident with the \textit{ab initio} theoretical prediction, and from these data, we can fit the $ g $-factor close to 2 for this spin state.
	
	\begin{figure}
		\centering
		\includegraphics[width=0.45\textwidth]{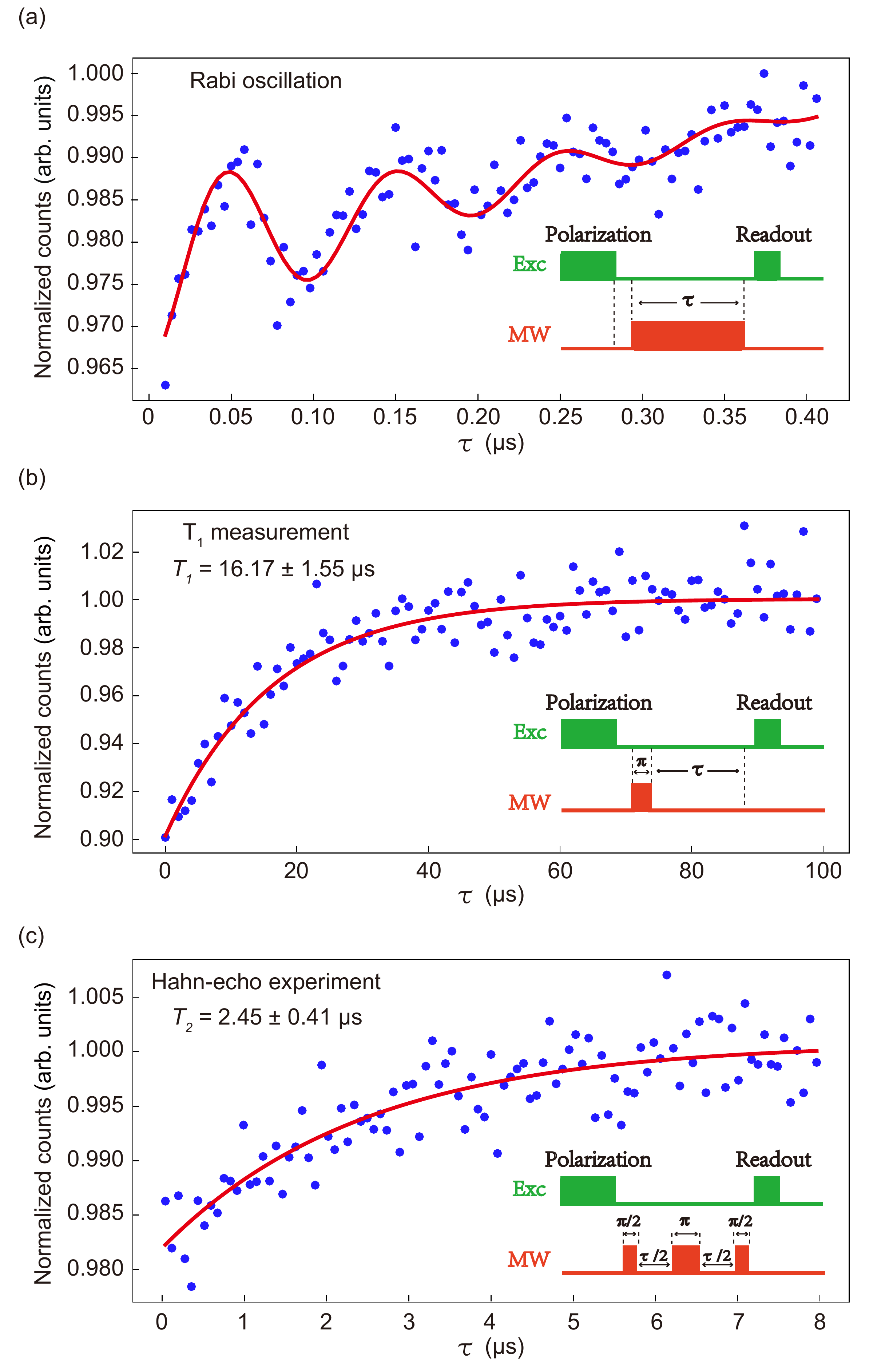}
		\caption{\label{Fig4} \textbf{Room-temperature coherent spin manipulation of Defect A.} (a) Rabi oscillation of Defect A. The inset schematically shows the pulse sequence, in which the first excitation laser pulse is used for the initialization of the spin state and the second pulse is used for a readout of the state after the MW pulse. (b) Spin-lattice relaxation time ($ T_{1} $) measurement of Defect A. The inset schematically shows the pulse sequence, in which the duration of the MW $ \pi $-pulse is obtained from Rabi oscillation and the waiting time $ \tau $ is varied. We can obtain the fitted $ T_{1}=16.17\pm 1.55$ $  \mu $s. (c) Hahn-echo experiment of Defect A. The inset schematically shows the pulse sequence. We can obtain the fitted $ T_{2}=2.45\pm 0.41$ $  \mu $s. All measurements are performed under a 35.8-mT magnetic field at room temperature.}
	\end{figure}
	
	\textbf{Room-temperature coherent control.} Next, we demonstrate the room-temperature coherent control of the single spin of Defect A in hBN. The pulsed ODMR is applied at the 35.8-mT magnetic field to manipulate the spin state, and the results are shown in Fig. 4. We first polarize the spin state to the ground state by optical excitation and then apply a MW pulse with variable lengths whose frequency is resonant with the spin-level splitting, followed by another laser pulse for the readout of the spin state. This is the process of Rabi-oscillation measurement, and the pulse sequence and result of Defect A are shown in Fig. 4a. This result is fitted by the Equation $ A+Be^{-\tau /T_{2}^{*}}\cos(2\pi f\tau+\phi)+Ce^{-\tau /T_{0}} $. The spin-dephasing time $ T_{2}^{*} $ obtained from the exponential decay of the oscillation term is $ \sim 140 $ ns. To determine the spin relaxation times $ T_{1} $ and $ T_{2} $ of Defect A, we perform $ T_{1} $ measurements and Hahn-echo experiments. The pulse sequences are shown as insets in Figs. 4b and 4c, where the parameters of $ \pi $-pulse and $ \pi /2 $-pulse are obtained from the results of Rabi oscillation. The results of the spin-lattice relaxation time $ T_{1} $ and spin-spin relaxation time $ T_{2} $ are shown in the corresponding main figures, where $ T_{1} $ and $ T_{2} $ are derived as $ \sim $16.17 $ \mu$s and $ \sim $2.45 $ \mu$s, respectively.
	
	\begin{table*}[htbp]
		\centering
		\setlength{\tabcolsep}{1mm}
		\renewcommand\arraystretch{1.2}
		\caption{The results of comparing our defects with reported spin defects. $ g^{(2)}(0) $ is the second-order autocorrelation measurement, ZPL is zero-phonon line, DOP is the degree of polarization.}
		\begin{tabular}{c c c c c c c c c}
			\hline \hline & & & & & & & & \\
			& $ g^{(2)}(0) $ & \tabincell{c}{Saturated\\ counts} & ZPL & \tabincell{c}{DW \\ factor} & DOP & \tabincell{c}{Coherent\\ control} & \tabincell{c}{$ T_{1} $ \\ (room / cryogenic)} & \tabincell{c}{$ T_{2} $ \\ (room / cryogenic)}  \\
			& & & & & & & &\\
			\hline & & & & & & & &\\
			V$ _{\text{B}}^{-} $ (ensemble) & $ - $ & $ - $ & $ - $ & $ - $ & $ - $ & Yes & 18 $ \mu $s / 12.5 ms \cite{gottscholl2021} & 2 $ \mu $s / 2 $ \mu $s \cite{gottscholl2021} \\
			& & & & & & & &\\
			\hline & & & & & & & &\\
			\tabincell{c}{Mendelson \emph{et al.} \cite{mendelson2021}\\ (ensemble)} & $ - $ & $ - $ & 585 nm & $ - $ & $ - $ & $ - $ & $ - $ / $ - $ & $ - $ / $ - $ \\
			& & & & & & & &\\
			\hline & & & & & & & &\\
			Stern \emph{et al.}  \cite{stern2021room} & 0.34 & 3.8$ \times 10^{4}$ s$^{-1} $ & $ \sim $ 580 nm & $ - $ & $ - $ & $ - $ & $ - $ / $ - $ & $ - $ / $ - $ \\
			& & & & & & & &\\
			\hline \tabincell{c}{\\ \\ Chejanovsky \\ \emph{et al.} \cite{chejanovsky2021}} & 0.59 & $ - $ & 720 nm & $ - $ & $ - $ & $ - $ & $ - $ / 17 $ \mu $s & $ - $ / $ - $\\
			& 0.22 & $ - $ & 760 nm & $ - $ & $ - $ & $ - $ & $ - $ / 13 $ \mu $s & $ - $ / $ - $  \\
			& & & & & & & &\\
			\hline & & & & & & & &\\
			\textbf{Guo \emph{et al.}$ ^* $} & \textbf{0.25} & \tabincell{c}{$ \mathbf{3.7\times 10^{6}} $ $ \mathbf{s^{-1}} $\\$ \mathbf{2.5\times 10^{7}} $ $ \mathbf{s^{-1}} $ \\ \textbf{(corrected)}} & $ \mathbf{\sim} $ \textbf{540} $ \mathbf{nm} $ & \textbf{0.8} & \textbf{0.8} & \textbf{Yes} & \tabincell{c}{$ \mathbf{16.17 \mu} $s /\\ \textbf{not detected}} & \tabincell{c}{$ \mathbf{2.45 \mu} $s /\\ \textbf{not detected}} \\
			& & & & & & & &\\
			\hline \hline
			$ ^* $Results in current work.
		\end{tabular}
	\end{table*}
	
	\textbf{Discussion}
	
	We compare Defect A with the other spin defects reported in hBN, as shown in Table I. The ultrabright and photostable single spin Defect A with the ZPL at $ \sim 546$ nm is significantly different from the other reported spin defects and exhibits the highest brightness and DW factor among these defects. The excellent optical properties can lead to high-efficiency optical initiation and readout of the spin with a high signal-to-noise ratio and are helpful for the further realization of the spin-photon interface based on the two-dimensional van der Waals material of hBN. The spin-relaxation times $T_1 \sim 16.17$ $ \mu$s and $T_2 \sim 2.45$ $ \mu$s of Defect A are comparable to those of the V$ _{\text{B}}^{-} $ ensemble at room temperature \cite{gottscholl2021}, which has the longest $ T_2 $ time of hBN among those reported. The room-temperature optical and spin properties, including the coherent control of the single spin defect in hBN, indicate that it can indeed be potentially applied in practical quantum-information applications in the future.
	
	\textbf{Theoretical simulation.} To explore the structure of Defect A, we vary the magnetic field, laser power and microwave power to perform ODMR measurements (see Supplementary Note 3). Fig. 5 shows the dependence of the ODMR linewidth on the magnetic field, laser power and microwave power, indicating that the linewidth has almost no dependence on the magnetic field and exhibits a power broadening effect. Regrettably, no obvious hyperfine structure can be observed even at very low laser power and microwave power, which might indicate relatively strong internuclear interactions among neighboring boron and nitrogen atoms mediated by the electron wavefunction. We apply low-power lasers and microwaves to decrease the power broadening effect and reduce nuclear spin polarization. The observed linewidth is approximately 22 MHz (Fig. 5b with 30-$ \mu $W laser and 0-dBm MW powers). Experimentally, it is a challenge to identify the atomic structure of defect emitters. Theoretical models can be developed using density functional theory calculations, and the simulated key parameters, including ZPL, PSB, and ODMR broadening caused by the hyperfine interaction between the electron spin and nuclear spins, can be directly compared to experimental values to identify defect. A previous investigation assigned carbon-substituted boron (C$_\text{B}$) as a single defect spin with a FWHM of the ODMR spectrum of $\sim 43$~MHz~\cite{auburger2021towards}. However, the calculated ZPL of C$_\text{B}$ is 731 nm (1.695~eV) which does not coincide with the observed ZPL of Defect A ($ \sim $546 nm), even considering the anticipated inaccuracy of 0.2~eV in the applied method~\cite{auburger2021towards}. Considering the similar ODMR spectra of that ODMR center~\cite{chejanovsky2021} and Defect A in our study, our working model is a complex consisting of C$_\text{B}$ defects in which the spin density should be localized on C$_\text{B}$ with an $S=1/2$ spin state. The relatively narrow ODMR linewidth in the experiment is related to the spin density localization on either nuclear spin-free isotopes with high natural abundance or nuclear spin isotopes with small gyromagnetic ratio, which is followed by the spin density overlapping with neighboring boron and nitrogen atoms with nonzero nuclear spins. The likely impurity candidates are carbon, oxygen and silicon~\cite{li2022identification,smart2021intersystem}. Among the possible structures, the positively charged C$_\text{B}$O$_\text{N}$ pair defect with C$_\text{2v}$ symmetry could be a reasonable model. The calculated ZPL of C$_\text{B}$O$_\text{N}$ is 538 nm (2.303 eV), and the Huang-Rhys (HR) factor $S$ is 2.58. We further calculate the hyperfine tensor of nuclear spins proximate to the C$_\text{B}$O$_\text{N}$ to simulate the broadening of the ODMR signal under varying external magnetic fields. A detailed discussion can be found in Supplementary Note 7. The peak position agrees well with the experimental result (Fig. 3b). The FWHM is approximately 23 MHz, slightly larger than the observed FWHM at low laser and MW powers when considering $^{11}$B isotopes. However, the C$_\text{2v}$ structure is not the most stable, and we observe a C$_\text{1h}$ symmetry configuration with carbon atoms moving out-of-plane. As a consequence, the ODMR linewidth increases to 50~MHz due to the localized wavefunction partially transferring to nitrogen atoms interacting with their nuclear spins. In addition, the C$_\text{B}$ and O$_\text{N}$ defects are both donors~\cite{weston2018native} in hBN so that the C$_\text{B}$O$_\text{N}$ complex has a high formation energy due to repulsion. Based on these findings, we further seek possible defects with a narrow ODMR linewidth in the stable ground state \cite{li2022carbon,li2022ultraviolet}. One facile treatment is to substitute the nearest boron (nitrogen) atoms close to C$_\text{N}$ (C$_\text{B}$), resulting in a carbon cluster, which may then result in a planar geometric configuration. C$_\text{2}$C$_\text{B}$ has an ODMR width of approximately 27 MHz, but the ZPL is much less than the current value~\cite{jara2021first,auburger2021towards}. Here, we propose a C$_\text{N}$C$_\text{B3}$ defect~\cite{berseneva2011mechanisms,maciaszek2022thermodynamics} in which the C$_\text{N}$ is surrounded by three C$_\text{B}$ atoms, as shown in Fig. 5(d). The positively charged C$_\text{N}$C$_\text{B3}$ shows a ZPL at 675.4 nm (1.836 eV) with $S$ = 1.51 (see Supplementary Note 7 for the C$_\text{N}$C$_\text{B3}$ details). The ODMR linewidth at 28~MHz in the electronic ground state implies that this defect configuration is a part of defect A. We further note that interlayer impurity oxygen interstitial is also considered (see Supplementary Note 7). The interlayer interstitial atom attracts atoms from the top and bottom layers, resulting in a distorted geometry of the $sp3$ bonding configuration. This again leads to a relatively large hyperfine interaction with neighbor boron and nitrogen atoms; therefore, it is likely that defect A is a planar defect similar to C$_\text{N}$C$_\text{B3}$. Although the exact structure of the defect is not identified here, our results strongly imply that the configurations of the substituted carbon and oxygen impurities play the key rule, where the spin density is mostly localized on the $p_z$ orbital of carbon.
	
	\begin{figure}
		\centering
		\includegraphics[width=0.5\textwidth]{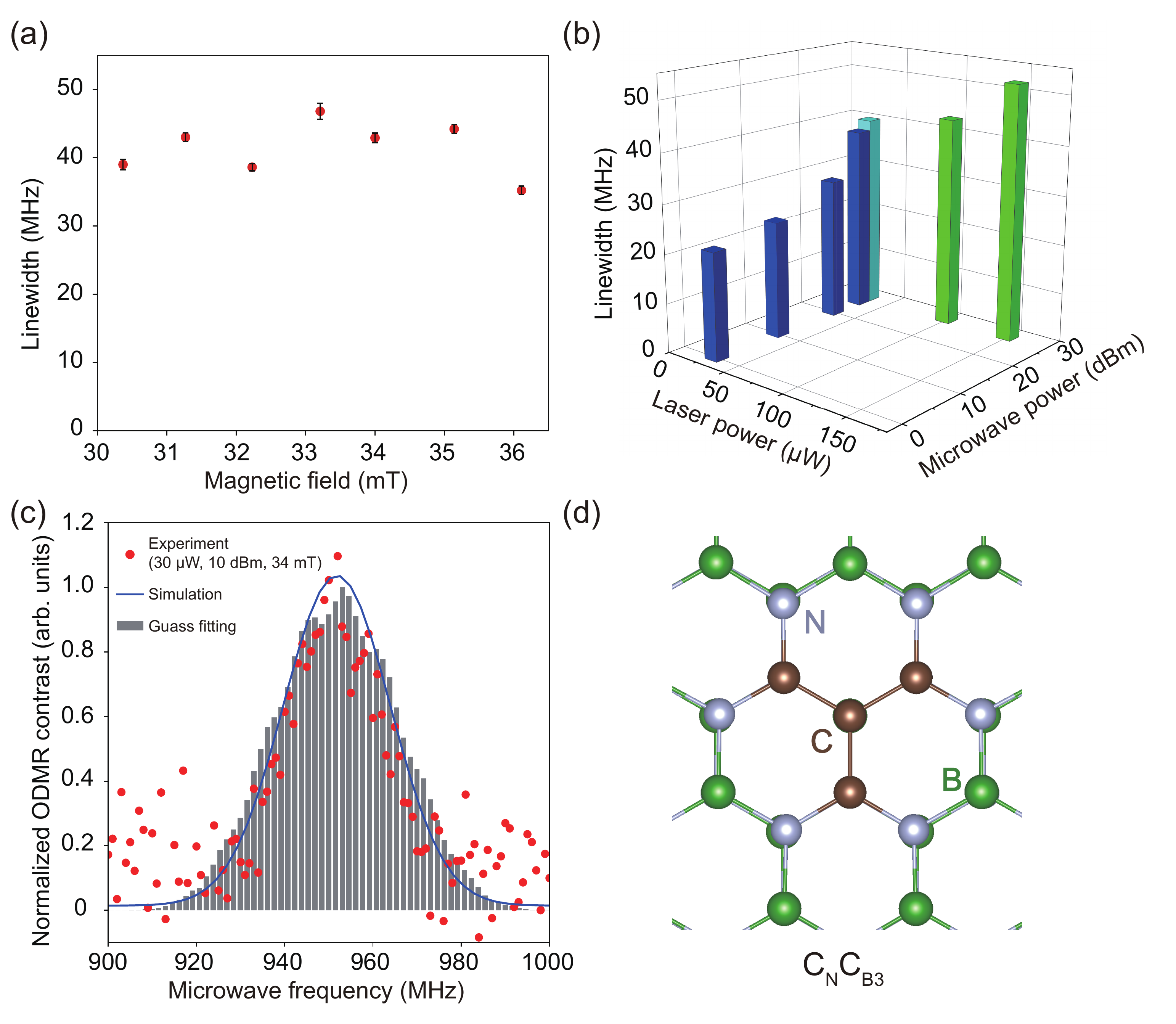}
		\caption{\label{Fig5} \textbf{ODMR linewidth analysis for Defect A.} (a) Dependence of the ODMR linewidth on the magnetic field. The error bars correspond to the fitting errors. The measurements are performed at 100-$ \mu $W laser power and 25-dBm microwave power at room temperature. (b) Dependence of the ODMR linewidth on laser power and microwave power. The measurements are performed under a 34-mT magnetic field at room temperature. (c) Comparison between the simulated and experimental ODMR signals with C$ _{\text{N}} $C$ _{\text{B3}} $ model. (d) Schematic structure of the defect C$ _{\text{N}} $C$ _{\text{B3}} $ in hBN.}
	\end{figure}
	
	\begin{table*}[htbp]
		\centering
		\setlength{\tabcolsep}{3.5mm}
		\renewcommand\arraystretch{1}
		\caption{Statistics on the probability of finding a single spin defect. The total defects in our work are screened out by the brightness threshold. P$ _{1} $ is the ratio of isolated defects with ODMR signal to the total defects investigated, P$ _{2} $ is the ratio of isolated defects with ODMR signal to all the isolated defects, and P$ _{3} $ is the ratio of isolated defects with evident Rabi oscillations to the isolated defects with ODMR signal.}
		\begin{tabular}{c c c c c c c c c}
			\hline \hline & & & & & & \\
			& \tabincell{c}{total defects \\ investigated} & isolated defects & \tabincell{c}{isolated defects with \\ ODMR signal} & P$ _{1} $ & P$ _{2} $ & P$ _{3} $\\
			& & &\\
			\hline & & & & & &\\
			Stern \emph{et al.}  \cite{stern2021room} & $ - $ & more than 400 & 27 & $ - $ & $ \sim $ 5\% & $ - $\\
			& & &\\
			\hline & & & & & &\\
			spin-coated sample & 50 & 13 & 2 & 4\% & 15.4\% & \multirow{4}{*}{10.5\%}\\
			& & & & & &\\
			\cline{1-6} & & & &\\
			isolated-flake-array sample & 20 & 20 & 17 & 85\% & 85\%\\
			& & & & & \\
			\hline \hline
		\end{tabular}
	\end{table*}
	
	To determine the overall properties of the spin defects in the hBN-defect array, we also measure the other color-center defects. Our method includes a brightness threshold (here, it is chosen as 0.5 MHz under 100-$ \mu $W laser excitation), which excludes the dark emission centers directly. We detect 20 emission points that are beyond the brightness threshold in the hBN array. All of their emissions exhibit an antibunching effect, and 17 of them exhibit a measurable ODMR signal with an external longitudinal magnetic field, suggesting a high yield of 85\%. As the comparison, we investigate the probability of finding such defects in directly spin-coated samples using the same brightness threshold. Fifty bright points are investigated, and only 2 of them are found to exhibit both an antibunching effect and an ODMR signal. The probability is 4\%. Detailed data are shown in Table II. Therefore, our method of creating an isolated-hBN-flake array and setting a brightness threshold enhances the success probability by 21-fold. We also perform Rabi oscillation measurements for all the isolated defects found with ODMR, and only 2 are revealed to have distinct Rabi oscillations, displaying a relatively low probability of demonstrating coherent control ($ \sim $ 10.5\%). Considering this low probability, the 21-fold enhancement induced by our method is critical to observing the coherent control of a single spin at room temperature.
	
	In Supplementary Note 4, we present the measurement results of some of the bright spots in the array sample. Most of these defects exhibit similar optical properties (ZPL at 540 $ \pm $ 10 nm) and spin signals (a positive ODMR contrast and the same position of the ODMR peak at the same magnetic field). Therefore, we speculate that they are the same type of defect, which is intrinsic in the hBN powder with a relatively low probability but can be found effectively using our method. During the measurements, we also find some other phenomena; for example, we observe the hyperfine structure of a single spin defect and a relatively high ODMR contrast ($ \sim $ 7\%) of another defect, which may be contributed to other new found defects (see Supplementary Note 6).
	
	In conclusion, we report and characterize a new type of ultrabright single spin defect in hBN and demonstrate the room-temperature quantum control of the hBN single spin, owing to our method of creating an isolated-hBN-flake array and setting a brightness threshold, which enhances the success probability of finding an isolated defect with an ODMR signal by 21-fold. The spin defects of hBN exhibit excellent quantum emission properties and significant coherent-spin characteristics. This defect may be a complex of carbon and oxygen dopants according to our \textit{ab initio} calculations. The room-temperature quantum control of the single spin in hBN combined with its good optical properties is a crucial step toward utilizing two-dimensional-material spin in quantum-information applications.

	\textbf{Methods}
	
	\textbf{Sample preparation.} A silicon wafer with 280-nm thermally-grown SiO$ _{2} $ on top is first cleaned by ultrasonication with acetone, isopropanol and deionized water in sequence. Then the pattern of the MW waveguide is formed on the wafer coated with positive S1813 resists by photolithography. A 100-nm gold film is coated on the wafer by an e-beam evaporator, and the extra gold is stripped by heating in N-Methylpyrrolidone (NMP) to form the gold film MW waveguide. Subsequently, 300-nm positive PMMA resist is spun onto the wafer and baked. Arrays of circular holes 520-nm in diameter are patterned into the PMMA on the gold film using 100-keV electron beam lithography. HBN ultrafine powders with 99.0\% purity \cite{Chen2021solvent} and $ \sim $70 nm particle size are obtained commercially from Graphene Supermarket and annealed in high vacuum at 1000$ ^{\circ} $C for 2 hours. A small amount of hBN nanopowder is mixed in ultrapure water and sonicated into a suspension. A drop of suspension is placed on the edge of the wafer, and then the slide attached to the wafer is slowly moved to disperse the droplets into the holes. Finally, the PMMA resists are dissolved in acetone, leaving arrays of hBN on the gold-film MW waveguide.
	
	\textbf{Fitting of optical propriety data.} The experimental $ g^{(2)}(\tau) $ data can be fitted well using the three-level Model $ g^{(2)}(\tau) = 1 - (1+a)e^{-|\tau|/\tau _1} + ae^{-|\tau|/\tau_2} $, where $a$ is a fitting parameter and $ \tau_1$ and $\tau_2$ are the lifetimes of excited and metastable states, respectively \cite{Tran2016Quantum}. This model indicates the existence of a metastable state, which is essential for room-temperature spin polarization and optical detection of spin resonance \cite{Widmann2015Coherent}. The radiative lifetime $\tau_1$$ \sim $2.5 ns is consistent with that previously reported, and the nonradiative lifetime $\tau_2$ is much longer ($\sim$600 ns) at 100-$ \mu $W laser power. The emission polarization is fitted by $ e_{1}+e_{2}\cos ^{2}(\theta+e_{3}) $, where $ e_1 $, $ e_2 $ and $ e_3 $ are the fitting parameters, and $ \theta $ is the polarization angle of the emitted fluorescence. The DOP ($ \sim $ 0.8) is derived by $ e_2/(2e_1+e_2) $. The experimental data of PL count versus excitation power are fitted by the Equation $ I=I_{\text{sat}}\times P/(P+P_{\text{sat}}) $, where $ I_{\text{sat}} $ is the saturated emission rate and $ P_{\text{sat}} $ is the saturation power. $ I_{\text{sat}}=3.7\pm 0.20 $ MHz and $ P_{\text{sat}}=446\pm 54 $ $\mu$W.

	\textbf{Theoretical simulation.} The \textit{ab initio} density functional theory (DFT) calculations were carried out within the Vienna \textit{ab initio} simulation package (VASP)~\cite{kresse1996efficiency, kresse1996efficient}. The projector augmented wave (PAW) formalism~\cite{blochl1994projector, kresse1999ultrasoft} is used to describe the electron and spin density near the nuclei. An $8\times8$ bulk hexagonal supercell with 256 atoms is constructed by embedding the point defects. The modified screened hybrid density functional of Heyd, Scuseria, and Ernzerhof (HSE)~\cite{heyd2003hybrid} is used to relax the structure and calculate the electronic properties. We increase the exact nonlocal Hartree-Fock exchange ($\alpha$ = 0.32) with the generalized gradient approximation of Perdew, Burke, and Ernzerhof (PBE)~\cite{perdew1996generalized}. The convergence criteria of total energy and quantum mechanical forces on the atoms are set to 10$^{-5}$ eV and 0.01~eV/\AA~with a plane wave cutoff energy of 450~eV. The Brillouin-zone is sampled by the single $\Gamma$-point scheme. The optimized interlayer distance was 3.37~\AA~with the DFT-D3 method of Grimme~\cite{grimme2010consistent} for dispersion correction. The excited states were calculated by the $\Delta$SCF method~\cite{gali2009theory,wu2005direct,alkauskas2014first}. The cw ODMR spectra can be calculated by the EASYSPIN toolbox based on the hyperfine matrix from DFT calculations \cite{stoll2006easyspin}. Here, the nuclear Zeeman and quadrupole interactions are not considered \cite{auburger2021towards}.
	
	\textbf{Data availability}
	
	The data that support the findings of this study are available within the article and its Supplementary Information. Additional relevant data are available from the corresponding authors upon reasonable request.
	
	\textbf{Code availability}
	
	The code used in this study is available from the corresponding authors upon reasonable request.

	\textbf{Acknowledgments}
	
	This work is supported by the Innovation Program for Quantum Science and Technology (No. 2021ZD0301200), the National Natural Science Foundation of China (Nos. 12174370, 11904356, 12174376, and 11821404), the Youth Innovation Promotion Association of Chinese Academy of Sciences (No. 2017492), the Open Research Projects of Zhejiang Lab (No.2021MB0AB02), the Fok Ying-Tong Education Foundation (No. 171007). This work was partially carried out at the USTC Center for Micro and Nanoscale Research and Fabrication. A.G. acknowledges the Hungarian NKFIH grant No. KKP129866 of the National Excellence Program of Quantum-coherent materials project and the support for the Quantum Information National Laboratory from the Ministry of Innovation and Technology of Hungary, and the EU H2020 project QuanTelCo (Grant No.\ 862721).
	
	\textbf{Author Contributions Statement}
	
	N.-J.G., W.L., Y.-T.W., J.-S.T., Y.-Z.Y. and X.-D.Z. conceived the experiments and analyzed the data with discussion of S.Y., Y.M., Q.L., J.-F.W., J.-S.X., R.-C. G. and G.-C.G.. N.-J.G., Y.-Z.Y. and X.-D.Z. performed the experiments with assistance of Z.-P.L., Z.-A.W. and L.-K.X.. S.L. and A.G. conceived and carried out the theoretical simulations. N.-J.G. W.L., Y.-T.W., J.-S.T., and S.L. wrote the manuscript. C.-F.L., A.G., and J.-S.T. supervised and coordinated the project.
	
	\textbf{Competing Interests Statement}
	
	The authors declare no competing interests.

\end{document}